\documentclass{jetplm}
\twocolumn

\usepackage{hyperref}

\lat

\title{On the collisions between particles in the vicinity\\
of rotating black holes\,\thanks{
On particles collisions in the vicinity of rotating black holes,
Pis'ma v ZhETF, vol.~{92}, iss.~3, (2010) 147--151.}}

\rtitle{\uppercase{\rm On the collisions between particles}}

\sodtitle{On particles collisions in the vicinity of rotating black
holes}

\author{\bf A.\,A.\,Grib$^{a}$, Yu.\,V.\,Pavlov$^{b}$}

\rauthor{\uppercase{\rm Grib, Pavlov}}

\address{$^a$Theoretical Physics and Astronomy Department, The
Herzen  University, 48, Moika, St.\,Petersburg, 191186, Russia\\
e-mail: andrei\_grib@mail.ru\\~\\
$^b$Institute of Problems in Mechanical Engineering, RAS, 61
Bolshoy, V.O., St.\,Petersburg, 199178, Russia\\
e-mail: yuri.pavlov@mail.ru}

\dates{9 June 2010}{*}

\abstract{Scattering of particles in the gravitational field of
rotating black holes is considered.
    It is shown that scattering energy of particles in the centre of mass
system can obtain very large values not only for extremal black
holes but also for nonextremal ones.
    Extraction of energy after the collision is investigated.
    It is shown that due to the Penrose process the energy of the particle
escaping the hole at infinity can be large.
    Contradictions in the problem of getting high energetic particles
escaping the black hole are resolved.}


\begin{document}

\maketitle

    In~\cite{GribPavlov2007AGN} we put the hypothesis that
active galactic nuclei can be the sources of ultrahigh energy
particles in cosmic rays observed recently by the AUGER group
(see~\cite{Auger07}) due to the processes of converting dark matter
formed by superheavy neutral particles into visible particles ---
quarks, leptons (neutrinos), photons.
    If active galactic nuclei are rotating black holes then
in~\cite{GribPavlov2007AGN} we discussed the idea that ``This black
hole acts as a cosmic supercollider in which superheavy particles of
dark matter are accelerated close to the horizon to the Grand
Unification energies and can be scattering in collisions.''
    It was also shown~\cite{GribPavlov2008KLGN} that in
Penrose process~\cite{Penrose69} dark matter particle can decay on
two particles, one with the negative energy, the other with the
positive one and particles of very high energy of the Grand
Unification order can escape the black hole.
    Then these particles due to interaction with photons close to
the black hole will loose energy analogously up to the
Greisen-Zatsepin-Kuzmin limit in cosmology~\cite{GZK1,GZK2}.

    First calculations of the scattering of particles in the ergosphere
of the rotating black hole, taking into account the Penrose process,
with the result that particles with high energy can escape the black
hole, were made in~\cite{PiranShahamKatz75,PiranShaham77}.
    Recently in~\cite{BanadosSilkWest09} it was shown that for the rotating
black hole (if it is the critical one) the energy of scattering is
unlimited.
    The result of~\cite{BanadosSilkWest09} was criticized
in~\cite{BertiCardosoGPS09,JacobsonSotiriou09} in the sense that it
does not occur in nature.
    The authors of~\cite{BertiCardosoGPS09,JacobsonSotiriou09} claimed that
if the black hole is not a critical rotating black hole so that its
dimensionless angular momentum $A \ne 1$ but $ A=0.998$ then the
energy is limited.

    In this paper we show that the energy of scattering in the centre of mass
system can be still unlimited in the cases of multiple scattering.
    In the \hyperref[2secBHColl]{first} part we calculate this energy, reproduce
the results
of~\cite{BanadosSilkWest09,BertiCardosoGPS09,JacobsonSotiriou09} and
show that in some cases (multiple scattering) the results
of~\cite{BertiCardosoGPS09,JacobsonSotiriou09} on the limitations of
the scattering energy for nonextremal black holes are not valid.

    In the \hyperref[3secBHColl]{second} part we obtain the results for the
extraction of the energy after collision in the field of the Kerr's
metric.
    It occurs that the Penrose process plays important role for getting
larger energies of particles at infinity.
    Our calculations show that the conclusion of~\cite{JacobsonSotiriou09}
about the impossibility of getting at infinity the energy larger
than the initial one in particle collisions close to the black hole
is wrong.

    The system of units $G=c=1$ is used in the paper.

\section{The Energy of Collision in the Field of Black Holes}
\label{2secBHColl}

    Let us consider particles falling on the rotating chargeless black hole.
    The Kerr's metric of the rotating black hole in Boyer--Lindquist
coordinates has the form                    
    \begin{eqnarray}
d s^2 = d t^2 -
\frac{2 M r \, ( d t - a \sin^2 \! \theta\, d \varphi )^2}{r^2 + a^2 \cos^2
\! \theta } \hspace{33pt}
\nonumber \\
-\, (r^2 \!+ a^2 \cos^2 \! \theta ) \left( \frac{d r^2}{\Delta} +
d \theta^2 \right)\! - (r^2 \!+ a^2) \sin^2 \! \theta\, d \varphi^2,
\label{Kerr}
\end{eqnarray}
    where
    \begin{equation} \label{Delta}
\Delta = r^2 - 2 M r + a^2,
\end{equation}
    $M$ is the mass of the black hole, $J=aM$ is angular momentum.
    In the case $a=0$ the metric~(\ref{Kerr}) describes the static chargeless
black hole in Schwarzschild coordinates.
    The event horizon for the Kerr's black hole corresponds to the value
    \begin{equation}
r = r_H \equiv M + \sqrt{M^2 - a^2} \,.
\label{Hor}
\end{equation}
    The Cauchy horizon is
    \begin{equation}
r = r_C \equiv M - \sqrt{M^2 - a^2} \,. \label{HorCau}
\end{equation}
    The surface called ``the static limit'' is defined by the expression
     \begin{equation}
r = r_0 \equiv M + \sqrt{M^2 - a^2 \cos^2 \! \theta} \,.
\label{PrSt}
\end{equation}
    The region of space-time between the horizon and the static limit is
ergosphere.

    For equatorial ($\theta=\pi/2$) geodesics in Kerr's metric~(\ref{Kerr}) one
obtains (\cite{Chandrasekhar}, \S\,61):
    \begin{equation} \label{geodKerr1}
\frac{d t}{d \tau} = \frac{1}{\Delta} \left[ \left(
r^2 + a^2 + \frac{2 M a^2}{r} \right) \varepsilon - \frac{2 M a}{r} L \right],
\end{equation}
    \begin{equation}
\frac{d \varphi}{d \tau} = \frac{1}{\Delta} \left[ \frac{2 M a}{r}\,
\varepsilon + \left( 1 - \frac{2 M}{r} \right)\! L \right],
\label{geodKerr2}
\end{equation}
    \begin{equation} \label{geodKerr3}
\left( \frac{d r}{d \tau} \right)^2 = \varepsilon^2 +
\frac{2 M}{r^3} \, (a \varepsilon - L)^2 +
\frac{a^2 \varepsilon^2 - L^2}{r^2} - \frac{\Delta}{r^2}\, \delta_1 ,
\end{equation}
    where
    $\delta_1 = 1 $ for timelike geodesics
($\delta_1 = 0 $ for isotropic geodesics),
$\tau$ is the proper time of the moving particle,
$\varepsilon={\rm const} $ is the specific energy:
the particle with rest mass~$m$ has the energy $\varepsilon m $ in the
gravitational field~(\ref{Kerr});
$ L m = {\rm const} $ is the angular momentum of the particle relative
to the axis orthogonal to the plane of movement.

    We denote~$x=r/M$, \ $ x_H=r_H/M$, \ $ x_C=r_C/M$, \ $ A=a/M$,
\ $ l_n=L_n/M$, \ $ \Delta_x = x^2 - 2 x + A^2 $.
    For the energy in the centre
of mass frame of two colliding particles with angular momenta $L_1, \, L_2$,
which are nonrelativistic at infinity ($\varepsilon_1 = \varepsilon_2 = 1$)
and are moving in Kerr's metric using~(\ref{Kerr}),
(\ref{geodKerr1})--(\ref{geodKerr3}) one obtains~\cite{BanadosSilkWest09}:
    \begin{eqnarray}
\frac{E_{\rm c.m.}^2}{2\, m^2} =
\frac{1}{x \Delta_x} \Biggl[ 2 x^2 (x-1) + l_1 l_2 (2-x)
\hspace{22pt}
\nonumber \\
+\, 2 A^2 (x+1) - 2 A (l_1 +l_2 ) \hspace{49pt}
 \label{KerrL1L2} \\
\hspace*{-4pt}
-\sqrt{\left( 2 x^2 + 2 (l_1 \!-\! A)^2 -l_1^2 x \right)
\left(2 x^2 + 2 (l_2 \!-\! A)^2 -l_2^2 x \right) } \Biggr].
\nonumber
\end{eqnarray}

    To find the limit $r \to r_H$ for the black hole with a given angular
momentum~$A$ one must take in~(\ref{KerrL1L2}) $x = x_H + \alpha$
with $\alpha \to 0 $ and do calculations up to the order~$\alpha^2$.
    Taking into account $ A^2 = x_H x_C$, $x_H + x_C=2$, after resolution
of uncertainties in the limit $\alpha \to 0 $ one obtains
    \begin{equation} \label{KerrLimA}
\frac{E_{\rm c.m.}(r \to r_H) }{2 m}
= \sqrt{1 + \frac{(l_1-l_2)^2}{2 x_C (l_1 -l_H) (l_2 - l_H)}} \,,
\end{equation}
    where $ l_H = 2 x_H/A $.

    For the extremal black hole $A=x_H=1$, $ l_H=2$ and
the expression~(\ref{KerrLimA}) is divergent when the dimensionless angular
momentum of one of the falling into the black hole particles $ l \to 2$.
    The scattering energy in the centre of mass system is increasing without
limit~\cite{BanadosSilkWest09}.

    Let's note, that to get the collision with infinite energy
one needs the infinite interval of as coordinate as proper time of
the free falling particle.
   Really, from Eqs.~(\ref{geodKerr1}), (\ref{geodKerr3}) for a particle with
dimensionless angular momentum~$l$ and specific energy $\varepsilon=1$
falling from some point $ r_0 = x_0 M $ to the point $r_f = x_f M > r_H$
one obtains for the coordinate time (proper time of the observer at rest
far from the black hole)
    \begin{equation} \label{KerrDelt}
\Delta t = M \! \int \limits_{x_f}^{x_0} \!\!
\frac{\sqrt{x} \left(x^3 +A^2 x + 2 A (A-l) \right)\, d x}{(x- x_H) (x- x_C)
\sqrt{ 2 x^2 - l^2 x + 2 (A -l)^2}} .
\end{equation}
    For the interval of proper time of the free falling to the black hole
particle one obtains from~(\ref{geodKerr3})
    \begin{equation} \label{KerrDeltau}
\Delta \tau = M \int \limits_{x_f}^{x_0}
\frac{x^{3/2}\, d x}{
\sqrt{ 2 x^2 - l^2 x + 2 (A -l)^2}} \,.
\end{equation}
     In extremal case ($A=1$, $l=2$)
the integrals~(\ref{KerrDelt}), (\ref{KerrDeltau})
diverges for $ x_f \to x_H=1 $ and
$ \Delta t \approx M 2 \sqrt{2} (x_f - 1)^{-1} $,
$ \Delta \tau \approx M | \ln(x_f - 1)| / \sqrt{2} $ for $ x_f \to 1 $.

    From~(\ref{geodKerr2}), (\ref{geodKerr3}) for the angle of the particle
falling in equatorial plane of the black hole one obtains
    \begin{equation} \label{KerrDelPhi}
\Delta \varphi = \int \limits_{x_f}^{x_0}\!\!
\frac{\sqrt{x} \left(x l + 2 (A-l) \right)\, d x}{(x- x_H) (x- x_C)
\sqrt{ 2 x^2 - l^2 x + 2 (A -l)^2}} .
\end{equation}
    If $A \ne 0$, then integral~(\ref{KerrDelPhi}) is divergent
for $ x_f \to x_H $.
    In extremal case ($A=1$, $l=2$)
$ \Delta \varphi \approx \sqrt{2} (x_f - 1)^{-1} $ for $ x_f \to 1 $.
    So before collision with infinitely large energy the particle must
commit infinitely large number of rotations around the black hole.

    Can one get the unlimited high energy of this scattering energy for
the case of nonextremal black hole?
    Formula~(\ref{geodKerr3}) leads to limitations on the possible values
of the angular momentum of falling particles:
the massive particle free falling in the black hole with dimensionless
angular momentum~$A$ being nonrelativistic at infinity ($\varepsilon = 1 $)
to achieve the horizon of the black hole must have angular momentum
from the interval
    \begin{equation} \label{geodKerr5}
- 2 \left( 1 + \sqrt{1+A}\, \right) =l_L \le l
\le l_R = 2 \left( 1 + \sqrt{1-A}\, \right).
\end{equation}

    Putting the limiting values of angular momenta $l_L, l_R$
into the formula~(\ref{KerrLimA}) one obtains the maximal values
of the collision energy of particles freely falling from infinity
    \begin{eqnarray} \label{KerrLimAMax}
E_{\rm c.m.}^{\, \rm inf}(r \to r_H) =
\frac{2m}{\sqrt[4]{1-A^2}} \hspace{44pt}
\\
\times \sqrt{\frac{1-A^2+\left( 1+ \sqrt{1+A} +\sqrt{1-A} \right)^2}
{1+\sqrt{1-A^2}} }\,.
\nonumber
\end{eqnarray}
    For $A=1-\epsilon$ with $\epsilon \to 0$ formula~(\ref{KerrLimAMax})
gives:
    \begin{equation} \label{KerrimAE}
E_{\rm c.m.}^{\, \rm inf}(r \to r_H)
\sim 2 \left( 2^{1/4}+2^{-1/4} \right) \frac{m}{\epsilon^{1/4}} \,.
\end{equation}
    So even for values close to the extremal $A=1$ of the rotating black hole
$E_{\rm c.m.}^{\, \rm inf}/ m$ can be not very large as mentioned
in~\cite{BertiCardosoGPS09,JacobsonSotiriou09}.
    So for $A_{\rm max} =0.998 $ considered as the maximal possible
dimensionless angular momentum of the astrophysical black holes
(see~\cite{Thorne74}), from~(\ref{KerrLimAMax}) one obtains
$ E_{\rm c.m.}^{\, \rm max} /m \approx 18.97 $.

    Does it mean that in real processes of particle scattering in
the vicinity of the rotating nonextremal black holes the scattering energy
is limited so that no Grand Unification or even Planckean energies can
be obtained?
    Let us show that the answer is no!
    If one takes into account the possibility of multiple scattering so that
the particle falling from infinity on the black hole with some fixed
angular momentum changes its momentum in the result of interaction with
particles in the accreting disc and after this is again scattering close to
the horizon then the scattering energy can be unlimited.

    The limiting value of the angular momentum of the particle close
to the horizon of the black hole can be obtained from the condition
of positive derivative in~(\ref{geodKerr1}) $dt /d \tau > 0$,
i.e. going ``forward'' in time.
    So close to the horizon one has the condition
$l < \varepsilon 2 x_H/A$ which for $ \varepsilon =1$
gives the limiting value $l_H$.

    From~(\ref{geodKerr3}) one can obtain the permitted interval in~$r$ for
particles with $ \varepsilon = 1 $ and angular momentum $l = l_H - \delta $.
    To do this one must put the left hand side of~(\ref{geodKerr3})
to zero and find the roots:
    \begin{equation} \label{KerrInttoch}
x_{1,2} = \frac{ l^2 \pm \sqrt{ l^4-16(A-l)^2} }{4} \,.
\end{equation}
    In the second order in~$\delta$ close to the horizon one obtains
    \begin{equation} \label{KerrIntl}
l = l_H - \delta \ \ \Rightarrow \ \ \
x \lesssim x_H + \frac{\delta^2 x_C^2}{4 x_H \sqrt{1-A^2} } \,.
\end{equation}
    The effective potential defined by the right hand side
of~(\ref{geodKerr3}) leads to the following behaviour of the particle.
    If the particle goes from infinity to the black hole it can achieve
the horizon if the inequality~(\ref{geodKerr5}) is valid.
    However the scattering energy in the centre of mass frame given
by~(\ref{KerrLimAMax}) is not large.
    But if the particle is going not from the infinity but
from some distance defined by~(\ref{KerrIntl}) then due to the form of
the potential it can have values of $l=l_H - \delta$
large than $l_R$ and fall on the horizon.
    If the particle falling from infinity with $ l \le l_R$ arrives
to the region defined by~(\ref{KerrIntl}) and here it interacts with other
particles of the accretion disc or it decays into a lighter particle which
gets an increased angular momentum $l_1 = l_H - \delta $,
then due to~(\ref{KerrLimA}) the scattering energy in the centre of mass
system is
    \begin{equation} \label{KerrInEn}
E_{\rm c.m.} \approx \frac{m}{\sqrt{\delta}} \,
\sqrt{ \frac{ 2(l_H - l_2) }{ 1- \sqrt{1 - A^2} }}
\end{equation}
    and it increases without limit for $\delta \to 0$.
    For $A_{\rm max} =0.998 $ and $l_2=l_L$, \
$ E_{\rm c.m.} \approx 3.85 m / \sqrt{\delta} $.  

    Note that for rapidly rotating black holes $A= 1 - \epsilon$
the difference between $l_H$ and $l_R$ is not large
    \begin{eqnarray}
l_H - l_R &=& 2 \frac{\sqrt{1-A}}{A} \left( \sqrt{1-A} + \sqrt{1+A} -A \right)
\nonumber \\
&\approx& 2 (\sqrt{2}-1) \sqrt{\epsilon}\,, \ \ \ \epsilon \to 0 .
\label{KerrInDLR}
\end{eqnarray}
    For $A_{\rm max} =0.998 $, \ $ l_H - l_R \approx 0.04$   
so the possibility of getting small additional angular momentum in
interaction close to the horizon seems much probable.
    The probability of multiple scattering in the accretion disc depends on
its particle density and is large for large density.

    Here we consider the model when the gravitation of the accretion disc
is treated as some perturbation much smaller than the gravitation of
the black hole so it is not taken into account.
    One must also mention that ``particles'' are considered as elementary
particles and not macroscopic bodies.
    So their ``large'' energy is limited by a Planckean value and we neglect
back reaction of it on the Kerr's metric of the macroscopic black hole.
    Electromagnetic and gravitational radiation of the particle surely
can change the picture but one needs exact calculations to see
what will be the balance.

\section{The Extraction of Energy after the Collision in Kerr's Metric}
\label{3secBHColl}

    Let us consider the case when there occurred for some $r>r_H$
interaction between particles with masses~$m$,
specific energies $\varepsilon_{1}$, $ \varepsilon_{2} $,
specific angular momenta $ L_{1}, \ L_{2} $
falling into a black hole,
so that two new particles with rest masses~$\mu$ appeared,
one of them~(1) moved outside the black hole, the other~(2)
moved inside it.
    Denote the specific energies of new particles as
$\varepsilon_{1 \mu}$, $ \varepsilon_{2 \mu} $, their angular momenta
(in units of $\mu$) as $ L_{1 \mu}, \ L_{2 \mu} $, \ $v^i = dx^i/ds $
 --- their 4-velocities.
    Consider particle movement in the equatorial plane of the rotating
black hole.

    Conservation laws in inelastic particle collisions for the energy
and momentum lead to
    \begin{equation} \label{I1}
m ( u_{(1)}^i + u_{(2)}^i ) = \mu ( v_{(1)}^i + v_{(2)}^i ) .
\end{equation}
    Equations~(\ref{I1}) for $t$ and $\varphi$-components can be written as
    \begin{equation} \label{I5K}
m (\varepsilon_1 + \varepsilon_2) =
\mu (\varepsilon_{1 \mu} + \varepsilon_{2 \mu}) ,
\end{equation}
    \begin{equation} \label{I5KL}
m (L_1 + L_2) = \mu (L_{1 \mu} + L_{2 \mu}) ,
\end{equation}
    i.e. the sum of energies and angular momenta of colliding particles is
conserved in the field of Kerr's black hole.
    The initial particles in our case were supposed to be nonrelativistic
at infinity: $ \varepsilon_{1} = \varepsilon_{2}=1 $ and therefore
for the $r$-component from~(\ref{geodKerr3}) one obtains
    \begin{eqnarray}
- m \left[ \sqrt{ \frac{2 M}{r^3} \,(a - L_1)^2 + \frac{2 M}{r} -
\frac{L_1^2}{r^2}} \right.
    \hspace{44pt} \nonumber  \\
+\left. \sqrt{ \frac{2 M}{r^3} \,(a - L_2)^2 +
\frac{2 M}{r} - \frac{L_2^2}{r^2}}\, \right]
    \hspace{37pt} \label{I4Ken11}
\\  \hspace{-8pt}
= \mu \left[ \sqrt{ \varepsilon_{1 \mu}^2
+\frac{2 M}{r^3} (a \varepsilon_{1 \mu} \!- L_{1 \mu})^2 +
\frac{a^2 \varepsilon_{1 \mu}^2 \!- L_{1 \mu}^2 - \Delta}{r^2} }
\right.
\nonumber  \\
-\left. \sqrt{ \varepsilon_{2 \mu}^2 +
\frac{2 M}{r^3} \, (a \varepsilon_{2 \mu} \!- L_{2 \mu})^2 +
\frac{a^2 \varepsilon_{2 \mu}^2 \!- L_{2 \mu}^2  - \Delta}{r^2} }
\, \right]\!.    \nonumber
\end{eqnarray}
    The signs in~(\ref{I4Ken11}) are put so that the initial particles and
the particle~(2) go inside the black hole while particle~(1) goes
outside the black hole.
    The values $ \varepsilon_{1 \mu}, \varepsilon_{2 \mu}$ are constants
on geodesics (\cite{Chandrasekhar}, \S\,61)
so the problem of evaluation of the energy at infinity extracted from
the black hole in collision reduces to a problem to find these values.

    For the case when the collision takes place on the horizon of the
black hole ($r \to r_H$) the system~(\ref{I5K})--(\ref{I4Ken11}) can be solved
exactly
    \begin{equation} \label{Inew}
\varepsilon_{1 \mu} \!=\! \frac{A L_{1 \mu}}{2r_H}, \
\varepsilon_{2 \mu} \!=\! \frac{2 m}{\mu} - \frac{A L_{1 \mu}}{2 r_H}, \
L_{2 \mu} \!=\! \frac{m}{\mu} (L_1+L_2) - L_{1 \mu} .
\end{equation}

    In general case the system of three Eqs.~(\ref{I5K})--(\ref{I4Ken11})
for four variables
$\varepsilon_{1 \mu}, \ \varepsilon_{2 \mu}, \ L_{1 \mu}, \ L_{2 \mu}$
can be solved numerically for a fixed value of one variable
(and fixed parameters $m/\mu, \ L_1/M, \ L_2/M, \ a/M, \ r/M$).
    The example of numerical solution is
$\mu/m=0.3, l_1=2.2, \, l_2=2.198, \, A=0.99, \, x=1.21, \, l_{1 \mu} = 16.35,
\, l_{2 \mu} = -1.69, \, \varepsilon_{1 \mu}=7.215 ,  
\, \varepsilon_{2 \mu}=-0.548 $.  
    Note that the energy of the second final particle is negative and the
energy of the first final particle contrary to the limit obtained
in~\cite{JacobsonSotiriou09} is larger than the energy of initial particles as
it must be in the case of a Penrose
process~\cite{PiranShahamKatz75,PiranShaham77}.
    What is the reason of this contradiction?
    Let us investigate the problem carefully.

    Note that if one neglects the states with negative energy in ergosphere
energy extracted in the considered process cannot be larger than
the initial energy of the pair of particles at infinity, i.e.~$2m$.
    The same limit~$2m$ for the extracted energy for any (including Penrose
process) scattering process in the vicinity of the black hole was
obtained in~\cite{JacobsonSotiriou09}.
    Let us show why this conclusion is incorrect.

    If the angular momentum of the falling particles is the same
then (see~(\ref{I4Ken11})) one has the situation similar to the usual
decay of the particle with mass~$2m$ in two particles with mass~$\mu$.
    Due to the Penrose process in ergosphere it is possible that the particle
falling inside the black hole has the negative relative to infinity energy
and then the extracted particle can have energy larger than~$2m$.

    The main assumption made in~\cite{JacobsonSotiriou09} is the supposition
of the collinearity of vectors of 4-momenta of the particles falling inside
and outside of the black hole in limiting case $A=1$, $l_1=2$
(see (9)--(11) in~\cite{JacobsonSotiriou09}).
    The authors of~\cite{JacobsonSotiriou09} say that these vectors are
``asymptotically tangent to the horizon generator''.

    In the limiting case
the expressions $d t / d \tau $,  $d \varphi / d \tau $
of the components of the 4-velocity of the infalling
particle~(\ref{geodKerr1}), (\ref{geodKerr2}) go to infinity when
$r \to r_H $, but $d r / d \tau $ goes to zero.
    In spite of smallness of $r$-components in the expression of the square
of the 4-momentum vector they have the factor $g_{rr} $ going to infinity
at the horizon.
    For example from~(\ref{Kerr}), (\ref{geodKerr3}) it is easy to see that
    \begin{equation} \label{22grr}
g_{rr}(x) u_{(1)}^r u_{(1)}^r \to -2 , \ \ \ \ x \to x_H .
\end{equation}
    So putting them to zero can lead to a mistake.
    To see if $u_{(1)}$ and $v_{(1)}$ are collinear it is necessary to put
the coordinate~$r$ of the collision point to the limit $r_H=M$ and resolve
the uncertainties $\infty / \infty$ and $0/0$.
    For the falling particle $\varepsilon =1$, $l=2$ the expressions for
components of the 4-vector~$u$ can be easily found
from~(\ref{geodKerr1})--(\ref{geodKerr3}).
    For the particle outgoing from the black hole due to exact solution
on the horizon~(\ref{Inew}) one puts
$\varepsilon_{1 \mu} = l_{1 \mu}/2 + \alpha$, where $\alpha$ is some function
of~$r$ and $l_{1 \mu}$, such that $\alpha \to 0$ when $r \to r_H$.
    Putting this $\varepsilon_{1 \mu} $ into~(\ref{geodKerr1})--(\ref{geodKerr3})
one gets for $x=r/M \to 1$
    \begin{equation} \label{Otn1}
\frac{v^t_{(1)}}{u^t_{(1)}} = \frac{v^\varphi_{(1)}}{u^\varphi_{(1)}} =
\frac{\alpha}{x-1} + \frac{l_{1 \mu}}{2} \,,
\end{equation}
    \begin{equation} \label{Otn2}
\frac{v^r_{(1)}}{u^r_{(1)}} = - \sqrt{
\frac{ 2 \alpha^2 }{(x-1)^2} + \frac{ 2 l_{1 \mu} \alpha}{x-1} +
\frac{3}{8}\, l_{1 \mu}^2 - \frac{1}{2} } \,.
\end{equation}
    Due to the condition $dt / d \tau >0 $ (movement forward in time)
the necessary condition for collinearity is that
both~(\ref{Otn1}) and (\ref{Otn2}) must be zero, which is not true.
    The notion of the asymptotic behaviour of 4-vectors on the horizon
(asymptotic collinearity) needs delicate mathematical analysis.
    One must find the limit of the norm of vectors.
    But this norm is defined by the scalar product of vector on itself
and it contains finite terms of the type~(\ref{22grr}) neglected by
the authors~\cite{JacobsonSotiriou09}, when they put the hypothesis
that the momentum of ejecta particle is~$\lambda k$.
    This hypothesis means that the $r$-components in~(\ref{Otn2})
are taken as identically zero and~(\ref{Otn1}) is enough for collinearity,
however~(\ref{22grr}) and~(\ref{Otn2}) show that this is incorrect.
    This leads to the conclusion that the considerations of the authors
of~\cite{JacobsonSotiriou09} for scattering exactly on the horizon can not be
used for the real situation of particle scattering close to the horizon.


\end{document}